\documentclass[conference]{IEEEtran}
\usepackage{blindtext, graphicx}
\usepackage{xcolor}
\usepackage{caption}

\ifCLASSINFOpdf
\else
\fi
\begin{document}
%
\title{Actors in VR storytelling}

\author{\IEEEauthorblockN{Selma Rizvic\IEEEauthorrefmark{1},
Dusanka Boskovic\IEEEauthorrefmark{1},
Fabio Bruno\IEEEauthorrefmark{2}, 
Barbara Davidde Petriaggi\IEEEauthorrefmark{3},
Sanda Sljivo\IEEEauthorrefmark{1} and
Marco Cozza\IEEEauthorrefmark{4}}
\IEEEauthorblockA{\IEEEauthorrefmark{1}Faculty of Electrical  Engineering\\
University of Sarajevo, Sarajevo, Bosnia and Herzegovina\\ 
Email: \{srizvic, dboskovic, sanda.sljivo \}@etf.unsa.ba}
\IEEEauthorblockA{\IEEEauthorrefmark{2}University of Calabria, Calabria, Italy\\
Email: fabio.bruno@unical.it}
\IEEEauthorblockA{\IEEEauthorrefmark{3}Istituto Superiore per la Conservazione ed il Restauro, Rome, Italy\\
Email: barbara.davidde@beniculturali.it}
\IEEEauthorblockA{\IEEEauthorrefmark{4}3D Research Srl, Cosenza, Italy\\
Email: marco.cozza@3dresearch.it}}


\maketitle

\begin{abstract}
Virtual Reality (VR) storytelling enhances the immersion of users into virtual environments (VE). Its use in virtual cultural heritage presentations helps the revival of the genius loci (the spirit of the place) of cultural monuments. This paper aims to show that the use of actors in VR storytelling adds to the quality of user experience and improves the edutainment value of virtual cultural heritage applications. We will describe the Baiae dry visit application which takes us to a time travel in the city considered by the Roman elite as "Little Rome (Pusilla  Roma)" and presently is only partially preserved under the sea.
\end{abstract}


%
\IEEEpeerreviewmaketitle

\section{Introduction}
Modern technologies offer various solutions for enhancing perception, awareness, and knowledge of cultural and natural heritage and thus contribute to its efficient conservation and sustainable promotion. Virtual, augmented and mixed reality (VR, AR, and MR) can add to our surroundings a component which cannot be seen in reality and revive the "genius loci" (the spirit of the place). Virtual reconstructions of decayed cultural monuments, or those who do not exist anymore or only their remains are preserved, lead our users through time travels and introduce them to important historical objects and characters.

The use of digital media platforms and interactivity for narrative purposes, either for fictional or non-fiction stories can be considered as interactive digital storytelling (IDS) \cite{Miller}. IDS enables the user to influence the flow and sometimes even the content of the story. The writers of interactive digital stories’ scenarios are adjusting their storytelling methodology to the Internet media. They have on disposal an enhanced set of tools, instead of pure text, to tell the story. In collaboration with the director and other team members, the writer decides on the narrative method, characters, and user interaction, to raise the attractiveness for the users.  

Motivation for this work is to add to immersion of users in VR environments emotions and empathy transferred through actors playing roles of characters inhabiting cultural monuments. 3D reconstructions without people do not convey accurately the information on the life in the past. Characters breathe life in virtual reconstructions of cultural heritage \cite{EurReview}. 

VR storytelling happens in 360-degree virtual environments for Head Mounted Displays. Because the user is in the center of the story, it is impossible to use the classical film language and shot composition tools to attract his/her attention to the plot. VR storytelling can be imagined as a theatre play with the audience placed in the middle of the stage. However, the 360-degree view reflects the way we perceive our natural surroundings, so, if properly designed, the VR storytelling could provide the full immersion of users in virtual worlds.

The adoption of immersive VR technologies for promoting underwater cultural heritage is being studied in the context of the iMARECulture Project \cite{iMARE} since it gives the possibility to simulate the diving experience and visit ancient shipwrecks and sunken cities. 

\begin{figure}[h]
    \centering
    \includegraphics[width=0.45\textwidth]{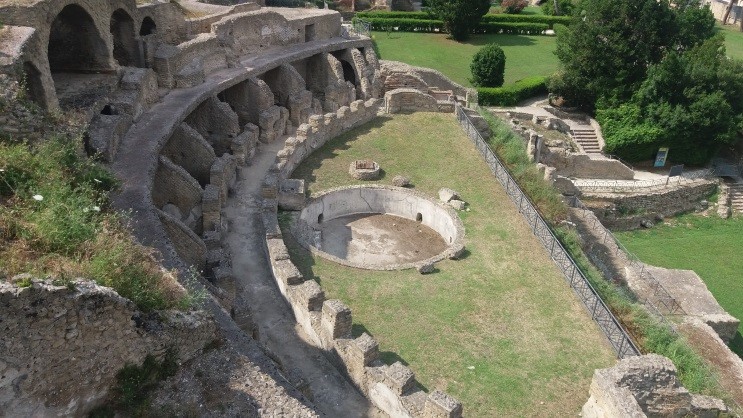}
    \caption{Baiae archeological park}
    \label{fig_baiae}
\end{figure}

In order to tell the story happening in the virtual reconstruction of a seaside villa in Baiae (Figure \ref{fig_baiae}), the city which was a luxurious resort for the Roman Empire aristocracy, we introduced the characters/actors and placed them inside the 360-degree video backgrounds. VR storytelling was combined with a VR simulation of diving inside the underwater city remains. This paper will show how the users perceive such a combination and how immersive, educational and entertaining is this kind of virtual cultural heritage presentation.

The paper is structured in the following way. Section 2 discusses similar projects with actors in 360-degree videos/animations, their advantages, and drawbacks in comparison with our method. Section 3 presents the Baiae Dry Visit application, its structure, and implementation overview, while Section 4 offers a detailed description of our VR storytelling methodology. In Section 5 we describe the user experience evaluation. Finally, in Section 6 we offer our conclusions.

\section{Related Work}
 A 5D modelling approach (3D geometry plus time plus levels of details) \cite{doulamis20155d} 
required for preservation and assessment of outdoor large scale cultural sites, was considered when creating the 3D model of Villa con Ingresso a Protiro from Baiae underwater excavation site, which we used as virtual environment for our storytelling scenes. 

Several similar projects consider actors in 360-degree videos or virtual environments. In this section, we describe some of the most relevant to our research. One of the first usages of actors in VEs was an adventure game Phantasmagoria \cite{Phan} by Sierra in 1995. Instead of the 3D model or 2D sprite, the main characters were actors filmed in front of a green screen with the pre-rendered VEs added as backgrounds. Alongside Phantasmagoria, other similar adventure games also used actors as main characters. At that time when 3D modeling was not as developed as today, this approach was a good solution for detailed graphics in the games.

On the other hand, games today have 3D or 2D sprites characters because of their high level of customization. There are some examples like The Quiet Man from Square Enix (2018) \cite{SqrEx} with a combination of live action video, in-game cutscenes and standard gameplay with 3D characters and environments. Unfortunately, the users reported a lack of immersion experience and the game did not have good reviews after release. Maybe if this game was developed as a VR game for HMD, the users could have better immersion experience. Besides games, actors in VEs are also used in digitalization of cultural heritage as in Livia's Villa Reloaded \cite{Liv}. In this project, actors are telling the story in the 3D reconstruction of the Livia's Villa. Livia was a Roman emperor Augustus' wife. The authors have developed three types of interactive digital storytelling application: Kinect application that uses gestures, online WebGL application and application for HMD. During the development, they have experimented with a combination of virtual reality paradigms, natural interaction interfaces, cinematographic techniques, and virtual set practices. The authors also had paid special attention to the user's orientation in the application using audio and visual cues. As a drawback, this project lacks user evaluation and could be improved with the evaluation of all three types of application.

Another project that preserves cultural heritage is Samuel Beckett in VR \cite{Beck}, where a team of interdisciplinary professionals, made an interactive VR narrative of Samuel Beckett’s theatrical text Play. They filmed the actors in front of a green screen using free-viewpoint video (FVV) and then added the VE as a background. Authors have decided to use FVV instead of 360-degree video, because of the 360-degree video limits the user’s viewpoint to the camera position imposed by the moviemaker. This project has also experimented with the organization of the author-audience dynamic. As Livia’s Villa, this project also lacks thorough user evaluation.

In Elmezeny \cite{Elm} the authors consider two dimensions of immersion in 360-degree videos: narrative and technical. They also did user evaluation with a comparison between 360-degree videos and traditional linear videos. As limitations to their approach, the authors state that 360-degree videos are still mostly driven by narratives. They have stated that cues are essential for directing the user attention in 360-degree videos. They also note that cues could be overstated if users just want to explore the environment, especially for videos which do not have narrative aspects. When it comes to videos without narrative aspects, there are a few examples of music videos done in this technology as in \cite{Scout, Drag} mostly as artistic experiments. It seems that these examples are just starting points and there is an excellent potential for 360-degree videos to become widespread storytelling methodology.

\section{Baiae dry visit application}
The Marine Protected Area-Underwater Park (MPA-UP) of Baiae was created on August 7th, 2002 by the Italian Ministry of Environment and Land Conservation, acting together with the Ministry of Cultural Heritage - now MIBAC, the Ministry of Transport, the Ministry of Agrarian Policies and with the Regione Campania \cite{Parks}.  

The MPA-UP is located off the north-western coasts of the bay of Puteoli (Naples), in the littoral zone between the southern limit of the port of Baiae and the dock of Lido Augusto. This site is part of the coastal region known as Campi Flegrei, that has been characterized by a periodic volcanic and hydrothermal activity, and it has been subjected to bradyseism, namely gradual changes in the levels of the coast concerning the sea level. Since antiquity, this coastal region has been subject to this phenomenon, which may be positive or negative, and in its present state, the remains of the Roman Era are submerged at a depth ranging between one and 14 to 15 meters below sea level.

Ancient Baiae was a bathing resort for the Roman aristocracy between the 2nd century BC and the 4th century AD. It was famous for its luxurious seaside villas, baths, shops, and coastal installations. In the late antiquity, Baiae began to sink into the water as a result of bradyseism and now a large part of the city is almost completely submerged. The MPA-UP, which has an area of about 176.6 hectares, not only safeguards the archaeological remains of the Roman city and infrastructures of the Roman harbor named \textit{Portus Iulius} but also represents an underwater area of great environmental value. Environmental aspects of this area are related to a peculiar volcanic and deformational history. Today the MPA-UP of Baiae has at least five itineraries open to the public, divers or no divers. This last type of tourists can visit the site on board of a boat with a transparent bottom. The local diving clubs are authorized by the Managing Authority - Parco Archeologico dei Campi Flegrei to accompany underwater tourist visitors to the submerged city. The underwater itineraries are: 1) the Nymphaeum of Punta dell'Epitaffio; 2) the Villa con ingresso a protiro - The Villa with a prothyrum entrance; 3) the Villa dei Pisoni; 4) Portus Iulius; 5) the "Secca fumosa."  

\begin{figure}[h]
    \centering
    \includegraphics[width=0.45\textwidth]{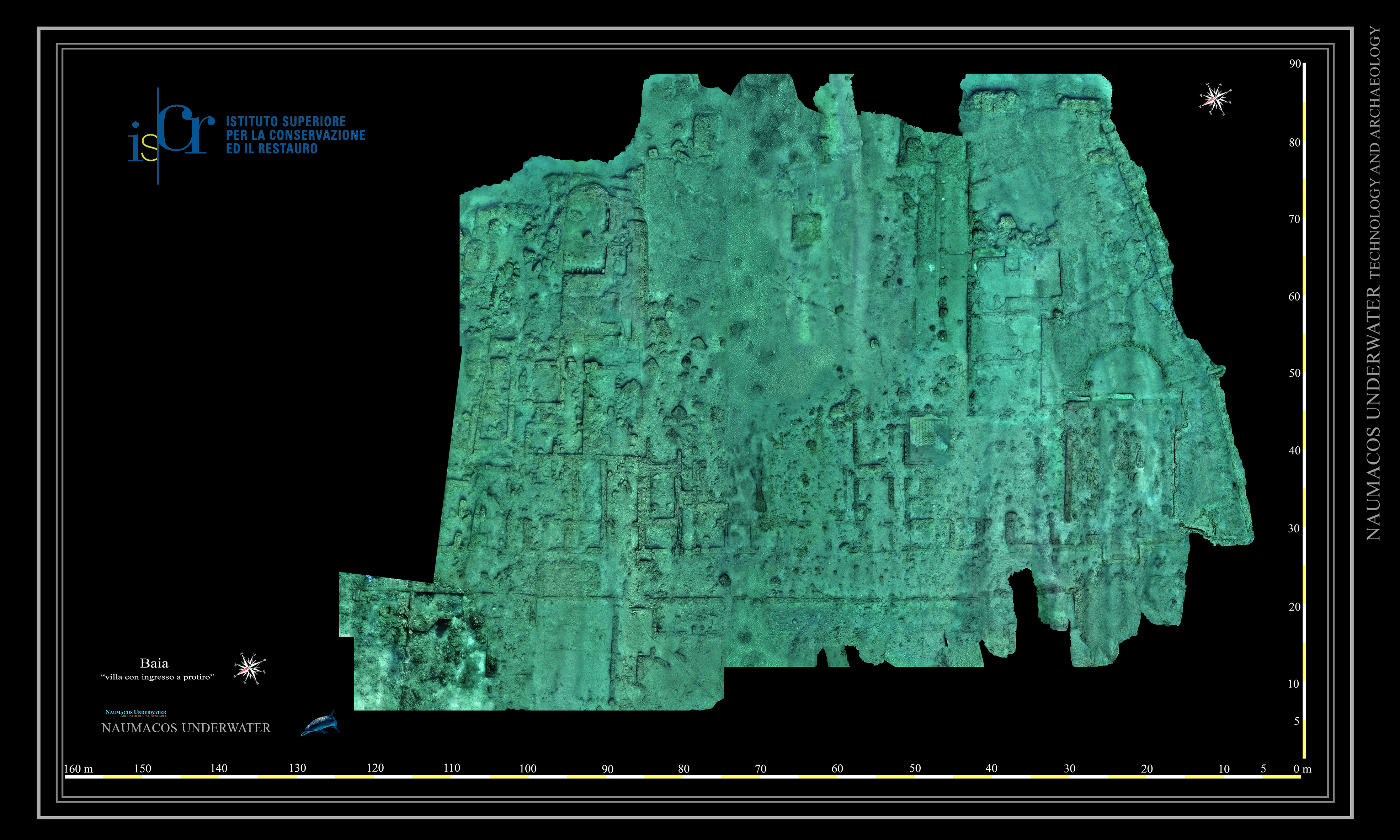}
    \caption{Ortophoto of the Villa con ingresso a protiro, \copyright ISCR}
    \label{fig_ortofoto}
\end{figure}

The purpose of the application was to show to the public the 3D reconstruction of the "Villa con ingresso a protiro," (Figure \ref{fig_ortofoto}) located at 5/6 meters depth, one of the many villae maritime scattered along the Lacus Baianus, and to increase the users’ knowledge of this underwater archaeological site.

The 3D visit of the villa is enriched by the history of its owner, \textit{Gaius Vibius Sabinus} (invented name), belonging to an important aristocratic family of Rome. He is involved in the wine trade overseas and owns the villa built just in front of the sea, along with the coast of Baiae, on the \textit{Lacus Baianus}, where he spends his time with his family and friends, relaxing. He is a cultured man, loving Greek art and philosophy; he is a collector of Greek masterpieces and well-done copies too. The "Villa con ingresso a protiro" has two clearly defined areas: the residential quarters and the baths (probably public).

Two red-plastered masonry benches - framed in stucco pilasters - mark out the threshold. The name 'prothyrum' comes from the presence of two stuccoed columns (no longer existing) that were edging two short parting walls built in front of the threshold. The vestibule, with doorways to the ostiarius quarters (\textit{Serapis} is the invented name in this history of the gatekeeper’s lodge), led to the atrium, whose walls were decorated with marble panels, as were those of adjacent areas, many of which had mosaic flooring. A mosaic floor in black and white tesserae with geometrical motifs (hexagons and pseudo-emblems with circles and peltae) is still visible in a room at the north-eastern corner of the atrium and is the room where \textit{Gaius Vibius Sabinus} receives \textit{Heliodorus} (invented name), an artist, sculptor, and copyist, famous for his opulent collection of copies of Greek masterpieces made by himself and his collaborators and his young assistant \textit{Saturninus} (invented name). The story takes place in different locations (the workshop of the sculptor \textit{Heliodorus}, the street with shops, the entrance of the villa, the atrium, the black and white mosaic paved room, the garden), adds to the quality of user experience and improves the edutainment value of this virtual cultural heritage application.   

\subsection{Application structure and implementation}

The Dry Visit application is a virtual exhibit that allows users to explore the 3D reconstruction of the Baiae underwater site using a Head Mounted Display (HMD). The HMD isolates the user from the distraction of the actual physical environment and encompasses the entire field of view, focusing the attention to the experience. The user navigates in the virtual environment simulating a real diving session from the scuba diving viewpoint, by moving his head and interacting with a single wireless handheld controller. In particular, to navigate, he presses the trigger of the handheld controller and points it in the direction they want to go. The use of VR to simulate the diving experience in underwater archaeological sites, and in particular on ancient shipwrecks, has demonstrated to be a useful educational tool able to engage people through a fantastic approach \cite{Bruno, Fotis}.

\begin{figure*}[h]
    \centering
    \captionsetup{justification=centering}
    \includegraphics[width=\textwidth]{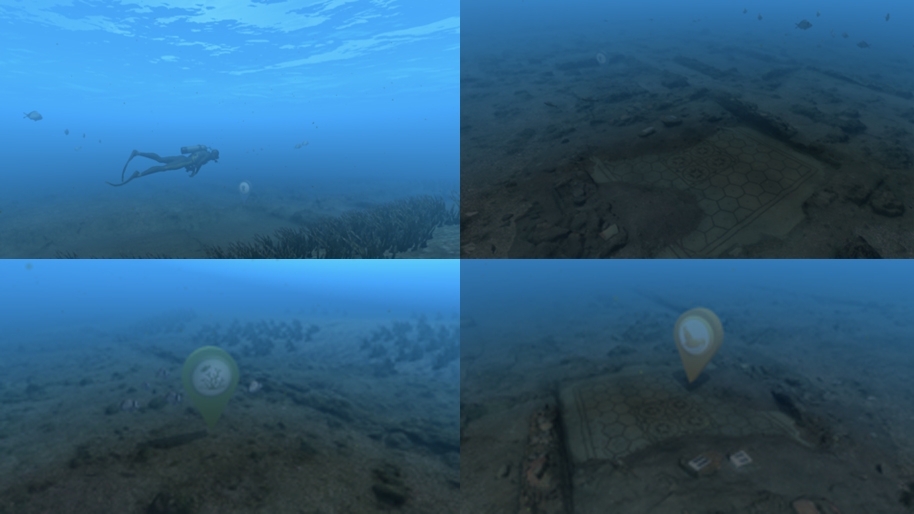}
    \caption{Underwater simulation screenshots}
    \label{fig_app}
\end{figure*}

The exploration of the underwater archaeological site starts above the water surface. In order to make a more attractive and engaging experience, the terrestrial environment representing the coastline that overlooks the archaeological site has been added to the scene. Moreover, the texturized 3D model representing the archeological site as it appears nowadays has been enhanced with the 3D models of flora and fauna. Also, several graphical effects, such as fogs, caustics, reflections, and refractions, have been combined in the scene in order to better simulate the underwater environment. Once the user dives in the submerged virtual environment, they are guided by a virtual diver among the storytelling videos (Figure \ref{fig_app}). The user can explore the area and play the videos in a precise order, by activating some particular objects placed in the virtual environment. When these objects are activated, the full-screen 360-degree video is played. Moreover, the user can see also the hypothetical reconstruction of the original structure of the Villa. The application is provided with a 3D model that represents the "Villa con ingresso a protiro" as it appeared in the past, and enables the user to explore the original status of the structures, environments and even furnishings. In particular, they can switch between the underwater environment and the hypothetical reconstruction by activating a particular object placed in the virtual environment.

The Dry Visit application has been developed using the Unity 3D game engine, and designed to provide usability for most type of users, implementing simple interactions and effective User Interfaces. More info on the implementation can be found in \cite{isprs-archives-XLII-2-W10-45-2019}.

\section{VR storytelling}
The main challenge in VR storytelling is that the user has control over what they see in the HMD. Because the users have freedom to turn around inside the 360-degree virtual environments, it is difficult to stage the story in such a way that they do not miss an essential part of the plot. All rules of shot composition that we learn in the film language grammar now have to change, as we cannot any more vary the shot size or make such shot composition to emphasize important elements in the scene. Some staging techniques can be borrowed from the theatre, but our audience is not in front of the stage, they are in its center. 

Guidelines we developed within the Sarajevo Charter \cite{RizvicVS} still apply for the VR interactive digital storytelling. Based on these guidelines we developed the Kyrenia application, recreating the story of one of the oldest shipwrecks found in the Mediterranean \cite{GCH17} through six stories with actors. After watching these stories, the user can embark on a virtual model of the ship. While in Kyrenia the stories have not been 360-degree videos, in the Mostar Cliff Diving VR project \cite{GCH18} we created 360-degree videos recording the various spots around the Old Bridge and added audio narration about the history of the bridge, its construction, destruction and reconstruction, as well as the interview with the cliff diving champion about that 300 years old tradition. After the users successfully test their knowledge gained from the stories, they can experience the virtual dive through our simulation. 

Although the results of the user experience evaluation show that the users felt immersion in both applications and qualified them with a high edutainment value, we consider that the drawback of Kyrenia project was the lack of 360-degree environments in the stories and of the Mostar project the lack of actors-narrators. Therefore, continuing our research with Baiae project, we implemented both and created VR stories with actors.

\subsection{The scenario}

In the pre-production stage of the Baiae VR storytelling project, we prepared the scenario, developed the visual styling and planned the production. The archaeological partners in the project provided the scenario with the story about a wealthy aristocrat purchasing a statue from an artist to decorate his garden. The scenario idea is based on statues remains found in the archaeological sites where gardens of luxurious villas were. Our story is happening in the “Villa con ingresso a protiro,” 3D reconstructed by the partner institution \cite{Davidde18}.

The storytelling conceived and written by Barbara Davidde Petriaggi and Roberto Petriaggi consists of 6 parts: the intro story (360-degree video of Baiae remains on land with a voice over introducing the viewer with the city and its historical significance), the sculptor's workshop where all characters introduce themselves (the sculptor, his apprentice, the aristocrat, and the slave), the street with shops (where the sculptor tells his apprentice how beautiful is Baiae), the villa entrance, the room with mosaics where the slave announces the visitors to the aristocrat, the atrium (the sculptor is introducing his apprentice with the villa) and the discussion of the sculptor and the aristocrat about the statue design and price.

\begin{figure}[h]
    \centering
    \includegraphics[width=0.45\textwidth]{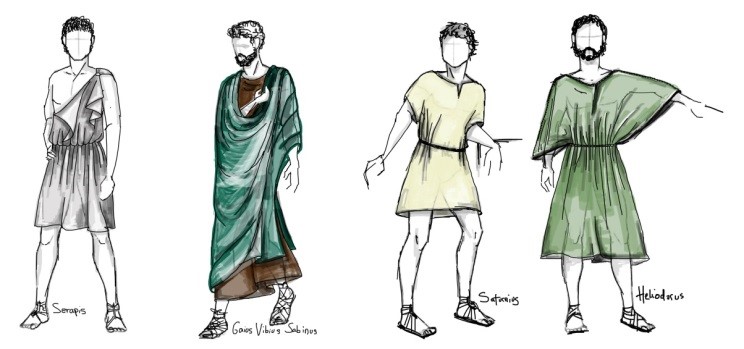}
    \caption{Costumes and visual styling of characters}
    \label{fig_costumes}
\end{figure}

The visual styling of actors and their costumes were created in collaboration with the archaeologists to accurately recreate the life in the Roman period. Costumes sketches are shown in Figure \ref{fig_costumes}.

According to the scenario and advice from the archaeological team, we made the actors casting. It was crucial to choose actors who look like Romans and who are of appropriate age for playing the characters in the stories.

Apart from the 3D model of the villa, we prepared 3D models of the sculptor's workshop, the street with shops and the villa entrance outer side, as backgrounds for storytelling. The models were created in 3ds max and rendered as 360-degree videos.

\begin{figure}[h]
    \centering
    \includegraphics[width=0.45\textwidth]{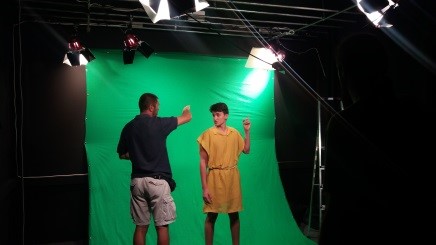}
    \caption{Filming the actors}
    \label{fig_actors}
\end{figure}

The actors were recorded against the green screen background (Figure \ref{fig_actors}) and superimposed in 360-degree video backgrounds in Adobe After Effects. After color correction and sound postproduction, the final 360-degree videos were exported. During preparations of backgrounds and filming, we have been previewing the scenes through Head Mounted Display, as the desktop appearance of 360-degree videos creates a significantly different impression of depth and view position.

\begin{figure*}[!t]
    \centering
    \includegraphics[width=\textwidth]{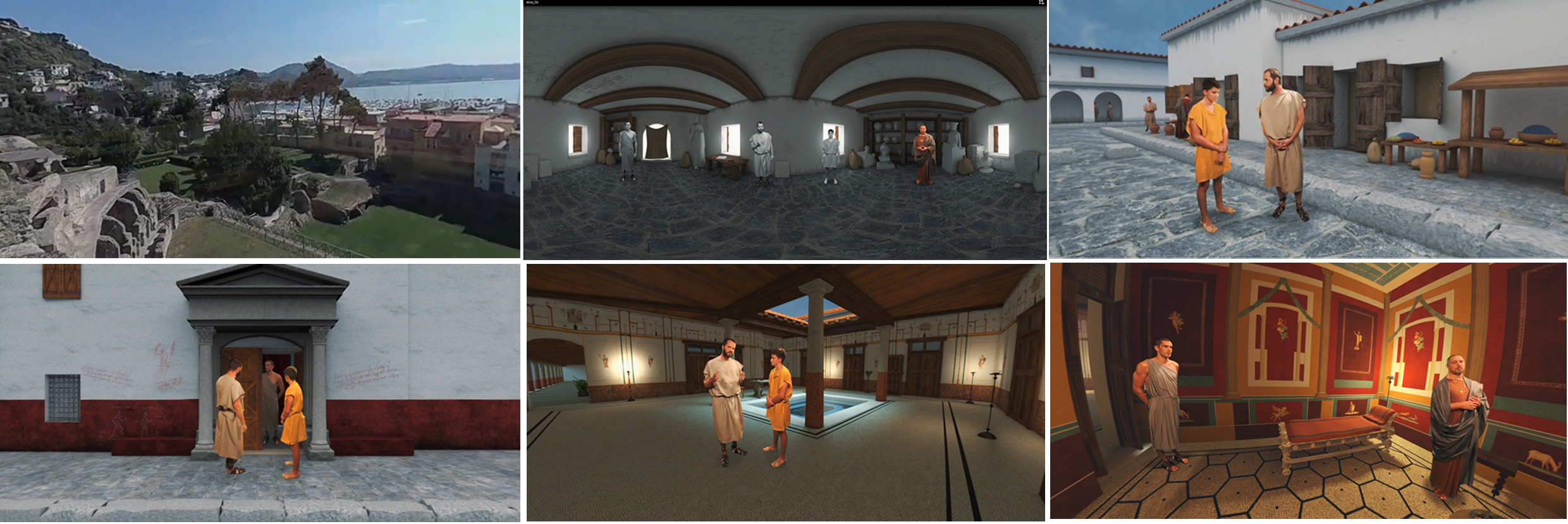}
    \caption{Screenshots from the VR stories}
    \label{fig_screen}
\end{figure*}

The final versions of stories were exported from Adobe After Effects in VR video H.264 format. The screenshots can be seen in Figure \ref{fig_screen}.

\section{User experience evaluation}
Our research question was "How the use of actors in the VR storytelling affects user experience related to the immersion and the edutainment value of virtual cultural heritage applications?" Building on our previous results in evaluating the user behaviors and the situational use of the VR technology \cite{GCH18} we have decided to conduct quantitative user experience study using structured questionnaire. The study was conducted in the research lab at the University and the evaluators were able to observe participants while engaged in the Baiae VR digital stories.

The participants were recruited by invitation, and we invited representatives of different user groups, aiming to balance user types regarding their age, their professional interests and their role in the educational process. In order to ensure that the responses represent diverse cross-section of respondents and to establish validity of survey responses, we included questions for relevant demographic data: age, sex, professional background, and their role in education, as recommended for self-selection of participants \cite{Lazar10}.

Our primary hypothesis was to show that the use of actors in VR storytelling adds to the quality of user experience and improves the immersion and the edutainment value of virtual cultural heritage applications.

\subsection{Experiment setup}

Before the commencement of the experiment participants were informed about the experiment procedure, their tasks and instructed to freely express their opinions. Immediately after engaging with the Baiae VR digital stories users answered the questions summarizing their perceived experience. 

Evaluators were observing the users during the experiment and noted user behavior, movements, objects of attention and possible undesirable effect of using the VR equipment. The complete experiment: interaction with the VR application and answering the web based post-interaction questionnaire lasted approximately 15 minutes.

\subsection{Participants}

The evaluation involved 23 participants. Summary of user demographics is as follows: professional background: STEM (n = 12), humanities (n = 6) and arts (n = 5); and their role in educational process: student (n = 12), teacher (n = 6) and other (n = 5). With respect to their age participants were categorized into following groups: younger than 20 (n = 2), ages 20-25 (n = 12), ages 25-35 years (n = 5), ages 35-50 years (n = 1), and aged older than 50 years (n = 3).

\subsection{Questionnaire}

The questionnaire contained three sections: (1) introductory part with demographics data, (2) main part with Likert scale items organized in 3 sub-scales measuring immersion, edutainment and ease of use, and (3) concluding section: user feedback and one question measuring effectiveness of educational part. 
Likert  items were defined as straightforward statements with positive logic and repetitions were avoided. Responses were delivered on a 5-point Likert range, with 1-strongly disagree and 5-strongly agree. The sub-scales were balanced comprising 7 items each.  

\subsection{Results}

Results are analyzed separately for each sub-scale presenting specific feature, namely: immersion, edutainment and easy of use. Summary statistical scores are presented in table \ref{tbl:stat1}.

\begin{table}[htb]
\caption {Summary results. } 
\label{tbl:stat1}
\centering
\begin{tabular} {lrr} 
\hline
Feature & Mean & Percent Agree \\ 
\hline
Immersion & 3.91 & 72\% \\
Edutainment & 4.22 & 81\% \\
Ease of use & 4.36 & 84\% \\     
\hline
\end{tabular}
\end{table}

List of Likert statements and their respective scores are provided in Table \ref{tbl:stat2}. Questions are split by sub-scale, with I.x addressing Immersion, E.x addressing edutainment and U.x  addressing Ease of Use.  Respective summary statistics measures are shown: Mean  and Percent Agree. Percent Agree corresponds to the proportion of users who responded with a view greater than neutral, towards agree.

\begin{table}[htb]
\caption {Detailed statistics}
\label{tbl:stat2}
\begin{tabular} {p{0.4cm}p{5cm}cp{0.7cm}} 
\hline
No  & Statement - item                          & Mean & Percent Agree \\ 
\hline
I1 & I found the 3D villa environment pleasant in terms of design and aesthetics. & 4.30 & 87\% \\
I2 & I found the actors interesting in terms of aesthetics. & 3.91 &  78\% \\
I3 & I found the voices of actors and narrator pleasant. & 3.52 &  61\% \\
I4 & I found the appearance of actors truthful and satisfactory. & 3.48 &  57\% \\
I5 & The VR story provides satisfaction. & 4.26 &  83\% \\
I6 & I have experienced the ancient Rome. & 3.83 &  70\% \\
I7 & Live actors make the VR story more personal and vivid. & 4.10 &  70\% \\
E1 & I found the VR story scenario clear and satisfactory & 3.96 &  74\% \\
E2 & I would like to use a similar VR story to learn about destroyed historical sites. & 4.61 &  87\% \\
E3 & I had fun watching the actors and viewing the Baiae surrounding. & 4.22 &  87\% \\
E4 & I think that the VR story respects the tradition of living in Roman city. & 4.00 &  74\% \\
E5 & I have learned new facts about living in the ancient Rome. & 3.95 &  70\% \\
E6 & The VR stories like Baiae can generate learning content and help the transmission and preservation of knowledge. & 4.61 &  91\% \\
E7 & The VR stories like Baiae should be included in the official learning processes. & 4.17 &  83\% \\
U1 & It is useful using $360^{o}$ view to get an overview of the scene. & 4.65 &  91\% \\
U2 & It is interesting being able to turn around and to look up the scene. & 4.83 &  96\% \\
U3 & I like position of the viewer in the center of the scene and among actors. & 3.87 &  70\% \\
U4 & The VR story innovate transmission of knowledge. & 4.70 &  91\% \\
U5 & Actors fit in the ancient Rome environment. & 3.83 &  74\% \\
U6 & I was comfortable with my role as a spectator. & 4.57 &  87\% \\
U7 & I followed the narrator and actors’ dialogues with ease. & 4.22 &  78\% \\
    \\ 
    \hline     
\end{tabular}
\end{table}
All three sub-scales contained several statements addressing explicitly the role of live actors, for example: I7, E3, U5.

It is significant to note that among the items with the highest score per sub-scale we can identify the most useful items as the following: "I1 - I found the 3D villa environment pleasant in terms of design and aesthetics," "I7 - Live actors make the VR story more personal and vivid," "E3 -I had fun watching the actors and viewing the Baiae surrounding," "U1 - It is useful using 360 view to get an overview of the scene," and "U2 It is interesting being able to turn around and to look up the scene."

In addition to the quantitative results we have found user observations facilitating analyses and providing for conclusions useful for our future work. Good example is the monitoring how the users reacted to the change of colour of the speaking actor as a visual cue to attract their attention, an important immersion factor in 360-degree VR environments \cite{Elm}. We have avoided a direct questioning but observed and noted that the directing user's attention had a positive response. We have noted that users' attention was directed to the speaking actor, regardless of their viewpoint, and the actor was easily recognized by his coloured appearance. The use of visual cues has particular importance for 360-degree videos that can not rely on framing and positioning the objects of interest in the center of user's visual attention within the frame.  

In our previous research we have identified the  difference in how students and teachers assess the benefits of bringing cultural heritage in the classroom using the VR \cite{JCLE19}. The following items in the edutainment sub-scale focus on the same questions: "E2 - I would like to use a similar VR story to learn about destroyed historical sites," and "E6 - The VR stories like Baiae can generate learning content and help the transmission and preservation of knowledge," with average scores: students 4.50 and teachers 4.91. It is important to note that in spite of a difference between the scores, both groups assessed the items with the highest score. 

\begin{figure}[h]
    \centering
    \includegraphics[width=0.45\textwidth]{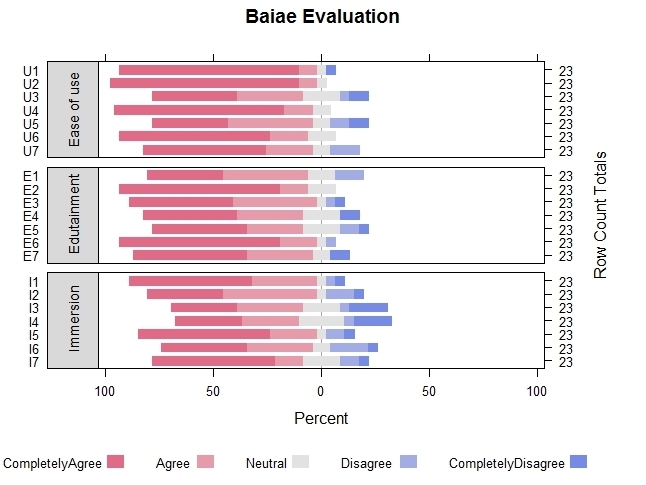}
    \caption{Visualization of Likert item analysis}
    \label{fig_likert}
\end{figure}

The graph presented in Figure \ref{fig_likert} provides insight into each specific item and illustrates distribution of responses grouped according to the sub-scales. Visualization is programmed in accordance with \cite{Robbins14}.

\subsection{Discussion} 

Summary results presented in Table \ref{tbl:stat1} and detailed item scores in Table \ref{tbl:stat2} proves our hypotheses that the use of actors in VR storytelling contributes to the quality of user experience and evokes positive level of immersion in virtual cultural heritage applications. 

Evaluation of edutainment was extended with a question aimed to measure effectiveness of educational dimension: "VR story helped me understand the social roles hierarchy in the ancient Rome." The average score of answers is 4.26. We have designed the statement to address implicit knowledge linked to the actors and their roles. The obtained results suggest that use of actors has a potential in education, users learn through transfer of experience and also through empathy with actors.    

The passive role as spectator was not assessed as negative. When asked for a feedback on "If I had an opportunity to act and influence the flow of the story I would" users were mainly interested in pausing the narrative and having more time for observing the environment, adding more details in the environment, being able to "play back." Only 2 of 23 participant expressed willingness to act as to provoke actors or to influence the flow of the story.

\section{Conclusion}

In this work, we explored whether the use of actors contributes to the immersion and edutainment of the VR  cultural heritage application. We placed the actors in virtual environments created as 3D reconstructions of various locations in the Baiae underwater excavation site. They were performing scenes from a scenario where a sculptor and his apprentice are visiting a rich Roman aristocrat, owner of the Villa con Ingresso a Protiro, in order to sell him a statue to decorate his garden. This storytelling is a part of a Dry visit application, aiming to enable the users to virtually dive using Head Mounted Displays and explore the underwater remains from Baiae archaeological park in Italy.

The achieved results are in accordance with our hypotheses.  The VR digital stories involving actors resulted in achieving a high level of user immersion. User scores for edutainment were even higher, and we can conclude that personalizing historical roles presents a novel and entertaining way for users to learn about historical sites and ancient societies.  Users’ feedback brings to our attention important feature for the future work: the ability to pause the story and explore the  360-degree environment and rewind the narrative to some point of interest. The interactivity could be extended to include a search for a specific object or to help the user to learn some important fact.

VR storytelling for cultural heritage still has a lot of research potential. In the future work we will focus on obtaining the right balance between narration and user interaction, to achieve maximum edutainment level. We will also work more on interaction methods within the virtual environments to prevent unpleasant effects of motion sickness and claustrophobia. We believe that in the future HMD devices will be accessible for everyday use and there will be a need for applications using their potential in communicating cultural heritage information to the general public.

\section*{Acknowledgment}

{
Research presented in this paper was supported by H2020 project 727153 iMARECULTURE. The Authors want to express warm thanks to the Parco Archeologico dei Campi Flegrei, for a kind cooperation with the iMARECULTURE project in Baiae, and to Roberto Petriaggi for his precious contribution to the VR Storytelling implementation.}  

\ifCLASSOPTIONcaptionsoff
  \newpage
\fi




\begin{thebibliography}{1}


\bibitem{Miller}
Carolyn Handler Miller. Digital storytelling: a creator’s guide to interactive entertainment. Elsevier/Focal Press, Amsterdam Boston, 2014.
  
\bibitem{iMARE}
Bruno F., Skarlatos D., Lagudi A., Agrafiotis P., Liarokapis F., Poullis C., Ritacco G., Cejka J., Kouril P., Philpin-Briscoe o., Mudur S., Simon B., Development and integration of digital technologies addressed to raise awareness and access to European underwater cultural heritage. An overview of the H2020 i-MARECULTURE project. Oceans'17 MTS/IEEE conference, Aberdeen, June 2017.  
  
\bibitem{Phan}
Daniel Kurlan, ‘Phantasmagoria,’ PC’s Best and Most Dated Horror Game, Turns 21, Deadpixels, https://bloody-disgusting.com/video-games/3403381/phantasmagoria-pcs-best-dated-horror-game-turns-21/ Visited on 1-2-2019

\bibitem{SqrEx}
Alice O'Connor, FMV brawler The Quiet Man out now, Rock paper shotgun, https://www.rockpapershotgun.com/2018/11/01/the-quiet-man-released/ Visited on 1-2-2019 

\bibitem{Liv}
E. Pietroni, M. Forlani and C. Rufa, "Livia's Villa Reloaded: An example of re-use and update of a pre-existing Virtual Museum, following a novel approach in storytelling inside virtual reality environments," \emph{2015 Digital Heritage}, Granada, 2015, pp. 511-518.

\bibitem{Beck}
O'Dwyer, Neill; Johnson, Nicholas; Bates, Enda; Pagés, Rafael; Amplianitis, Konstantinos; Monaghan, David; Smolic, Aljoscha. (2018). Samuel Beckett in Virtual Reality: Exploring narrative using free viewpoint video. Leonardo. 1-10.

\bibitem{Elm}
ELMEZENY, Ahmed; EDENHOFER, Nina; WIMMER, Jeffrey. Immersive Storytelling in 360-Degree Videos: An Analysis of Interplay Between Narrative and Technical Immersion. Journal For Virtual Worlds Research, [S.l.], v. 11, n. 1, Apr. 2018. ISSN 1941-8477

\bibitem{Scout}
Alannah Forman, VRScout https://vrscout.com/news/360-vr-music-videos-and-artists/, Visited on 2-2-2019

\bibitem{Drag}
https://www.youtube.com/watch?v=81fer9ulOeA, Visited on 2-2-2019 

\bibitem{Parks}
Ministero dei beni e delle attività culturali e del turismo, 2019
(http://www.parks.it/parco.sommerso.baia/par.php) Visited on 1-2-2019

\bibitem{Bruno}
Bruno F., Barbieri L., Lagudi A., Cozza, M., Cozza A., Peluso, R., Muzzupappa M., Virtual dives into the underwater archaeological treasures of South Italy. Virtual Reality, pp. 1-12, 15 June 2017.

\bibitem{Fotis}
Liarokapis, F., Kou{\v{r}}il, P., Agrafiotis, P., Demesticha, S., Chmel{\'{i}}k, J., \& Skarlatos, D. (2017). 3D modelling and mapping for virtual exploration of underwater archaeology assets. The International Archives of Photogrammetry, Remote Sensing and Spatial Information Sciences, 42, 425.


\bibitem{RizvicVS}	
S. Rizvic, N. Djapo, F. Alispahic, B. Hadzihalilovic, F. Fejzic-Cengic, A. Imamovic, D. Boskovic, V. Okanovic, Guidelines for interactive digital storytelling presentations of cultural heritage, In Proceedings of 9th International Conference on Virtual Worlds and Games for Serious Applications (VS-Games 2017), pp 1-7, ISBN 978-1-5090-5812-9 (Xplore)

\bibitem{GCH17}	
Selma Rizvic, Dusanka Boskovic, Vensada Okanovic, Sanda Sljivo, Kyrenia - Hyper Storytelling Pilot Application, In: Proceedings of 15th EUROGRAPHICS Workshop on Graphics and Cultural Heritage 2017, pp. 177–181.doi:10.2312/gch.20171311.

\bibitem{GCH18}	
E. Selmanovic, S. Rizvic, C. Harvey, D. Boskovic, V. Hulusic, M. Chahin and S. Sljivo, VR Video Storytelling for Intangible Cultural Heritage Preservation, In: Proceedings of 16th EUROGRAPHICS Workshop on Graphics and Cultural Heritage 2018

\bibitem{Davidde18}	
B. Davidde Petriaggi, R. Petriaggi, F. Bruno et alii, A digital reconstruction of the sunken “Villa con ingresso a protiro” in the underwater archaeological site of Baiae, IOP Conference Series: Materials Science and Engineering,June 2018,364 012013 OPEN ACCESS

\bibitem{Lazar10} 
LAZAR J., FENG J. H., HOCHHEISER H., Research Methods in Human-Computer Interaction, Wiley Publishing, 2010.

\bibitem{JCLE19}	
Rizvic, S., Boskovic, D., Okanovic, V. et al., Interactive digital storytelling: bringing cultural heritage in a classroom,J. Comput. Educ. (2019). https://doi.org/10.1007/s40692-018-0128-7

\bibitem{Robbins14}
HEIBERGER R., ROBBINS N.,  Design of diverging stacked bar
charts for likert scales and other applications, Journal of Statistical Software 57, 5 (Apr. 2014), 1–32

\bibitem{doulamis20155d}
Doulamis, Anastasios and Doulamis, Nikolaos and Ioannidis, Charalabos and Chrysouli, Christina and Grammalidis, Nikos and Dimitropoulos, Kosmas and Potsiou, Chryssy and Stathopoulou, Elisavet Konstantina and Ioannides, Marinos,  5D MODELLING: AN EFFICIENT APPROACH FOR CREATING SPATIOTEMPORAL PREDICTIVE 3D MAPS OF LARGE-SCALE CULTURAL RESOURCES, ISPRS Annals of Photogrammetry, Remote Sensing \& Spatial Information Sciences, 2015

\bibitem{isprs-archives-XLII-2-W10-45-2019}
Bruno, F. and Lagudi, A. and Barbieri, L. and Cozza, M. and Cozza, A. and Peluso, R. and Davidde Petriaggi, B. and Petriaggi, R. and Rizvic, S. and Skarlatos, D.,  VIRTUAL TOUR IN THE SUNKEN VILLA CON INGRESSO A PROTIRO WITHIN THE UNDERWATER ARCHAEOLOGICAL PARK OF BAIAE, ISPRS - International Archives of the Photogrammetry, Remote Sensing and Spatial Information Sciences, XLII-2/W10, 2019, 45--51

\bibitem{EurReview}
Rizvic, S.,How to Breathe Life into Cultural Heritage 3D Reconstructions, European Review, 2017, 25(1), pp. 39–-50

\end {thebibliography}
%

%

\begin{IEEEbiography}[{\includegraphics[width=1in,height=1.25in,clip,keepaspectratio]{picture}}]{John Doe}
\blindtext
\end{IEEEbiography}




\end{document}